\documentclass[authoryear,preprint,review,12pt]{elsarticle}

\graphicspath{{./figures/}}

\usepackage{float}

\usepackage[colorinlistoftodos]{todonotes}

\usepackage{amssymb,amsmath}
\usepackage{gensymb}

\usepackage{siunitx}

\usepackage{lineno}

\usepackage{url}
\usepackage{breakurl}
\usepackage[breaklinks]{hyperref}

\journal{Cold Regions Science and Technology}

\begin{document}

\begin{frontmatter}



\title{FlakeOut: A Geometric Approach to Remove Wind-Blown Snow from Terrestrial Laser Scans}


\author[inst1, cor1]{David Clemens-Sewall}
\ead{David.W.Clemens-Sewall.Th@dartmouth.edu}
\cortext[cor1]{Corresponding author}
\author[inst2]{Matthew Parno}
\author[inst1]{Don Perovich}
\author[inst1,inst3]{Chris Polashenski}
\author[inst1]{Ian A. Raphael}

\affiliation[inst1]{organization={Thayer School of Engineering at Dartmouth College},
            addressline={14 Engineering Dr.}, 
            city={Hanover},
            postcode={03755}, 
            state={NH},
            country={USA}}

\affiliation[inst2]{organization={Department of Mathematics, Dartmouth College},
            addressline={27 N. Main St.}, 
            city={Hanover},
            postcode={03755}, 
            state={NH},
            country={USA}}

\affiliation[inst3]{organization={US Army Corps of Engineers - Cold Regions Research and Engineering Laboratory - Alaska},
            addressline={Ft. Wainwright}, 
            city={Fairbanks},
            postcode={99703}, 
            state={AK},
            country={USA}}

\begin{abstract}
Wind-blown snow particles often contaminate Terrestrial Laser Scanning (TLS) data of snow covered terrain. However, common filtering techniques fail to filter wind-blown snow and incorrectly filter data from the true surface due to the spatial distribution of wind-blown snow and the TLS scanning geometry. We present FlakeOut, a filter designed specifically to filter wind-blown snowflakes from TLS data. A key aspect of FlakeOut is a low false positive rate of \num{2.8e-4}---an order of magnitude lower than standard filtering techniques---which greatly reduces the number of true ground points that are incorrectly removed. This low false positive rate makes FlakeOut appropriate for applications requiring quantitative measurements of the snow surface in light to moderate blowing snow conditions. Additionally, we provide mathematical and software tools to efficiently estimate the false positive rate of filters applied for the purpose of removing erroneous data points that occur very infrequently in a dataset.
\end{abstract}


\begin{highlights}
\item FlakeOut effectively filters wind-blown snow particles from TLS data with a false positive rate of just \num{2.8e-4}, an order of magnitude less than common techniques.
\item We provide tools to efficiently estimate the false positive rate of filters applied for the purpose of removing erroneous data points that occur very infrequently in a dataset.
\end{highlights}


\end{frontmatter}

\section{Introduction}
\label{sec:Introduction}

Accurate measurements of snow surface topography are critical for monitoring and modeling climate change on sea ice and tundra  \citep[e.g.][]{sturm_winter_2002,petrich_snow_2012,polashenski_mechanisms_2012,liston_distributed_2018,dery_observational_2009,sturm_characteristics_2001}. Understanding the wind-driven spatial redistribution of snow in particular requires many measurements of the snow surface to quantify the temporal evolution of snow thickness \citep{deems_fractal_2006,sturm_matthew_chapter_2009,trujillo_scaling_2009}. Considerable effort has been made to make manual measurements of snow thickness more efficient with devices such as the magnaprobe \citep{sturm_automatic_2018}. However, these data cannot directly produce maps or areal products of snow surface changes \citep{sturm_automatic_2018}. Terrestrial Laser Scanning (TLS, also known as LiDAR––-Light Detection and Ranging) is a line-of-sight, active remote sensing technique that can be used to generate three-dimensional point clouds of the snow surface \citep{deems_lidar_2013}. The sensor is a pulsed laser coupled with a detector and a mechanism (a rotating stage, mirror, or both) for reorienting the laser and detector between pulses. At each orientation, the laser emits a pulse and the detector records the two-way travel time of the reflection. The location of the points relative to the scanner is derived from the sensor orientation and the two-way travel time of each reflection. A review of LiDAR applied to snow thickness measurements can be found in \citet{deems_lidar_2013}.

\begin{figure}[H]
    \centering
    \includegraphics[width=13.5cm]{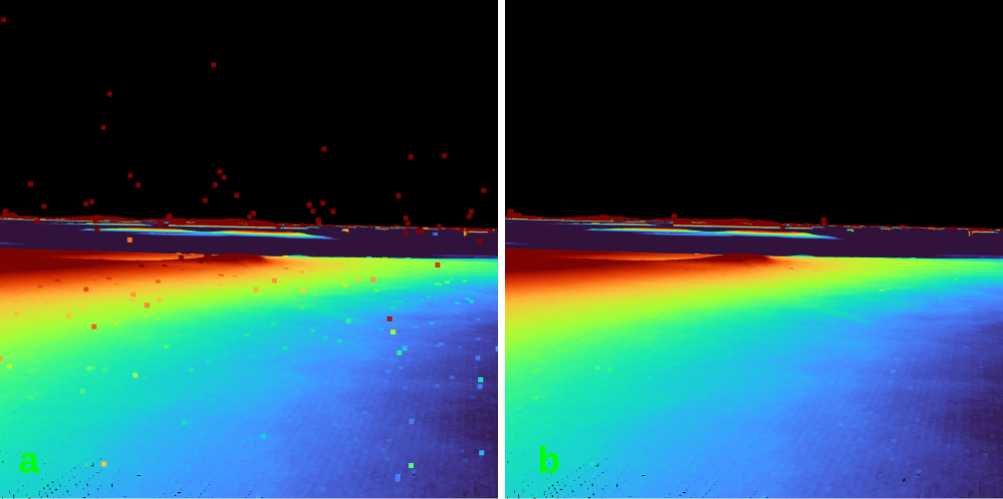}
    \caption{Rendering of TLS data collected on a day with mild blowing snow conditions before (a) and after (b) filtering out wind-blown snow particles using FlakeOut. Points are colored by vertical coordinate (blue is low, red is high) with a 10 cm color range. Most wind-blown snow particles are within 10 cm of the snow surface.}
    \label{fig:before_after}
\end{figure}

Open expanses like sea ice and tundra often experience dry snow and high wind speeds that lead to the frequent occurrence of blowing snow conditions \citep{li_probability_1997}. Fig. \ref{fig:before_after} shows a rendering of TLS data containing wind-blown snow particles. Blowing snow differs from falling snow in that the blowing particles are smaller than the original precipitation and are most frequently found in a saltating layer within approximately 10 cm of the snow surface \citep{schmidt_properties_1982,nishimura_blowing_2005}. Although researchers may be able to avoid making LiDAR measurements during active snowfall, LiDAR measurements in tundra and sea ice environments will likely need to contend with at least some blowing snow, which requires identification of blowing snow points in the TLS data. 

Multiple techniques exist for removing unwanted, isolated points (e.g. snow particles) from LiDAR point clouds. Many modern LiDAR detectors can record multiple returns per pulse if there are returns from objects that only partially occlude the pulse (e.g. snow particles). For these LiDAR systems, filtering all but the last return points---referred to herein as ``early return filtering''---is a common processing technique for removing above-ground returns \citep{contributors_pdal_2020}. Other common point cloud filtering routines include radius outlier removal and statistical outlier removal \citep{zhou_open3d_2018}. For radius outlier removal, all points that do not have at least a given number of neighbors within a given radius (both chosen by the user) are filtered out. Statistical outlier removal follows a similar principle; points are filtered by comparing their mean distance from neighboring points with the mean distance between neighboring points for the point cloud as a whole. While these techniques have seen success in many applications \citep[e.g.][]{zhou_open3d_2018}, the oblique scanning geometry of TLS and the low-lying nature of blowing snow particles make it particularly difficult for these techniques to filter blowing snow.

Two recent papers have proposed methods for filtering falling snow from mobile (vehicle-mounted) laser scanning data. \citet{charron_-noising_2018} proposed the dynamic radius outlier removal (DROR) filter. DROR operates under the same principle as radius outlier removal, except that the search radius around each point is determined by the range from the scanner to that point multiplied by the azimuthal resolution and a scalar. \citet{park_fast_2020} proposed the low-intensity outlier removal (LIOR) filter. LIOR identifies snow particles on the basis of the intensity of their returns being less than the intensity of returns from the surface. Experience with multiple TLS datasets for snow on sea ice (not shown) indicates that the return intensity varies considerably based on surface properties, snow particle orientation, and other conditions, so LIOR is not expected to be useful in this application. Both papers address only falling snow (as opposed to blowing snow) and are aimed at autonomous driving applications, not quantitative measurements of snow surfaces. Neither paper published their code, hence we do not apply these filters directly in this paper. However, we discuss differences between DROR and our proposed filter in Section \ref{sec:Discussion}.

In this paper we make two contributions. First, we propose a novel filter, FlakeOut (Section \ref{subsec:Filter}), that effectively filters wind-blown snow particles from TLS data while only incorrectly filtering 0.03\% of points from the true surface. To our knowledge, this is the first published filter directly addressing removing wind-blown snow particles from TLS data to enable quantitative measurements of the snow surface. Second, we provide the mathematical tools to estimate false positive rates for filtering rare events (in our data approximately one out of every thousand points is a wind-blown snow particle) without needing to manually classify hundreds of thousands or millions of points (Section \ref{subsec:Validation}). Additionally, we provide a custom visualization interface to efficiently manually classify our validation points. The source code for the FlakeOut filter, the visualization interface, and all analysis presented herein are publicly available \citep{clemens-sewall_davidclemenssewallflake_out_2021}.

\section{Materials and Methods}
\label{sec:MaterialsMethods}

\subsection{Data}
\label{subsec:Data}
We collected nine TLS scan positions with a Riegl VZ1000 on February 22, 2020 at approximately 88.6 \degree N, 55.6 \degree E on the Multidisciplinary Drifting Observatory for the Study of Arctic Climate \citep{shupe_mosaic_2020}. The ice topography of the scanned area included level ice, smooth multiyear ridges, and rough first year ridges. 2-3 cm of fresh snow had fallen February 18-21. Mild blowing snow conditions were observed at the time of the data collection. Operating parameters for the TLS are given in Table \ref{table:tls_param}. With these operating parameters, the stated maximum range of the VZ1000 on dry snow is 250 m. At each scan position, the TLS unit was mounted on a tripod approximately 2.5 m above the surface and recorded approximately 16.5 million data points. Human artifacts (snowmobiles, tents, etc) were manually removed from the data. Wind-blown snow particles are readily apparent in the data (see Fig. \ref{fig:before_after}).

\begin{table}[H]
\centering
 \begin{tabular}{|c c|} 
 \hline
 Parameter & Value \\ [0.5ex] 
 \hline
 Horizontal Scan Angle Range & 0\degree-360\degree \\
 Vertical Scan Angle Range & 30\degree-130\degree \\
 Horizontal Angle Stepwidth & 0.025\degree \\
 Vertical Angle Stepwidth & 0.025\degree \\
 Laser Pulse Repetition Rate & 300 kHz \\
 \hline
 \end{tabular}
\caption{Operating parameters for TLS scan positions}
\label{table:tls_param}
\end{table}

\subsection{FlakeOut}
\label{subsec:Filter}

FlakeOut, the filter presented here, consists of three stages. First, a simple vertical threshold filters all points that are higher than the highest snow or ice surface, usually a pressure ridge, in the scanned area. This is a convenience to filter points that are clearly above the snow and ice surface. The threshold value could be chosen by histogram analysis or manually. Second, the ``visible region'' filter uses the spatial relationship between the TLS scanner’s position and the last returns to determine the region of space visible to the scanner. Any early returns that are clearly inside this visible region (that is, the scanner sees empty space surrounding them), are classified as snow particles. Finally, the ``vertical z-score'' filter classifies points whose z-component is significantly higher than nearby points as snow particles.  Details of the visible region filter and z-score filters are provided below.

\subsubsection{Visible Region Filter}
\label{subsubsec:vis_region}

\begin{figure}[H]
    \centering
    \includegraphics[width=12cm]{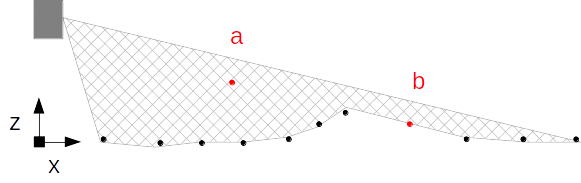}
    \caption{Schematic showing principle of operation of proposed filter. The gray rectangle in the upper left is the TLS scanner. Last returns are shown in black. The visible region is defined by the scanner and the last returns and is shown in the gray hashed area. Two early returns (a and b) are shown in red. Point a is within the visible area and would be classified as snow particle. Point b is on the edge of the visible area and would not be classified as a snow particle. Not to scale.}
    \label{fig:vis_region}
\end{figure}

The region of space that is visible to the scanner (the ``visible region'') is the region of space between the scanner and the surface defined by last returns (because, by definition, the scanner sees nothing past a last return). The visible region filter is implemented by, first, transforming all points into spherical coordinates with the origin at the scanner. Then, for each early return point, identifying all of the last return points that are adjacent (within the angular stepwidth of the scan multiplied by $\sqrt{2}$) in angular coordinates. If every adjacent point is further from the scanner than the early return point, then the early return point is strictly within the visible region and is classified as a snow particle. Otherwise, the early return point must be on the edge of the visible region, and is most likely be a partial return from the surface itself because surface points outnumber blowing snow particle points by approximately a factor of 1000. Fig. \ref{fig:vis_region} shows the principle of operation of the visible region filter. Although this filter cannot distinguish points that are actually snow particles but are on the edge of the visible region, such points are exceedingly rare and are not thought to have much of an impact on our ability to quantify the snow surface. The visible region filter is implemented in Python \citep{rossum_python_2010} using VTK \citep{schroeder_will_visualization_2006}. 

\subsubsection{Vertical Z-Score Filter}
\label{subsubsec:z_score}

The vertical z-score filter exploits the fact that, by definition, blowing snow particles are above the snow surface. First, the point cloud is separated into rectangular, laterally compact regions each containing a given number of points via a k-d tree decomposition implemented in SciPy \citep{virtanen_scipy_2020}. Then, for each region, the sample mean ($\bar{V}_r$) and standard deviation ($s_r$) of the vertical components ($v_{r,k}$) of the points in the region is computed. Here, $r$ is the region index and $k$ is a point index within that region. Next, compute the sample z-score ($z_{r,k}$) of each point in each region (Eqn. \ref{eqn:z}). Finally, classify all points whose sample z-score exceeds a given threshold as snow particles. Minimal testing suggests that regions of 100 points and a z-score threshold of 3.5 perform well.

\begin{equation}
\label{eqn:z}
z_{g,k} = \frac{v_{g,k} - \bar{V}_g}{s_g}
\end{equation}

\subsection{Validation}
\label{subsec:Validation}

Our filter is an automated, approximate binary classifier $\tilde{C}(x)$ that assigns a class $\tilde{y}\in\{S,G\}$ to each point in a point cloud on the basis of a set of features $x$ (e.g. elevation, etc...) characterizing each point. $S$ denotes that the point is a snow particle and $G$ denotes it is not a snow particle. We assess the accuracy of our approximate classifier by comparing it to manually classified points and computing false-positive and false-negative rates.  Here, we denote the manual classification of $x$ as $C(x)$. We employ false positive and false negative rates because they are intrinsic properties of the classifier and do not depend on the prevalence of snow particles in our particular dataset. We chose not to use the metric precision, which is commonly used \citep[e.g.][]{charron_-noising_2018}, because it depends on the prevalence of snow particles in a particular dataset and hence is less useful for comparing performance across datasets.

\subsubsection{Monte Carlo for False Positive and False Negative Rates}

The false positive rate is the probability ($\mathbb{P}$) that the approximate classifier $\tilde{C}(x)$ returns $S$ given that the true classifier $C(x)$ returns $G$.  Mathematically, this is given by

\begin{equation}
\label{eqn:fpr}
\mathbb{P}[(\tilde{C}(x)=S) \mid (C(x)=G)] = \frac{\mathbb{P}[(\tilde{C}(x) =S) \land (C(x)=G)]}{\mathbb{P}[C(x)=G]}
\end{equation}

Similarly, the false negative rate is the probability that the approximate classifier returns $G$ given that the true classifier returns $S$

\begin{equation}
\label{eqn:fnr}
\mathbb{P}[(\tilde{C}(x)=G) \mid (C(x)=S)] = \frac{\mathbb{P}[(\tilde{C}(x) =G) \land (C(x)=S)]}{\mathbb{P}[C(x)=S]}
\end{equation}

Computing these probabilities exactly would require manually classifying every point in the point cloud. This is clearly intractable given that LiDAR datasets contain hundreds of millions of points.  To overcome this issue, we have developed an importance sampling approach for efficiently constructing Monte Carlo estimates of the probabilities in \eqref{eqn:fpr}--\eqref{eqn:fnr}. The standard Monte Carlo estimate of $\mathbb{P}[(\tilde{C}(x) =S) \land (C(x)=G)]$ is:
\begin{eqnarray}
\mathbb{P}[(\tilde{C}(x) =S) \land (C(x)=G)] &=& \mathbb{E}_x\left[ I[ \tilde{C}(x)=S ]\, I[ C(x)=G ] \right]\\
&=& \frac{1}{K_{all}} \sum_{k=1}^{K_{all}} I[ \tilde{C}(x_k)=S ]\, I[ C(x_k)=G ]\\
&\approx& \frac{1}{K} \sum_{k=1}^K I[ \tilde{C}(x_k)=S ]\, I[ C(x_k)=G ]
\end{eqnarray}
where $k$ is the point index, $K_{all}$ is the total number of points in the point cloud, $K\ll K_{all}$ is a smaller number of points used to estimate the false positive rate, and $I[\cdot]$ is an indicator function ($1$ if True, $0$ if False). Following the same process leads to Monte Carlo estimates of the other probabilities

\begin{eqnarray}
\mathbb{P}[(\tilde{C}(x) =G) \land (C(x)=S)] &\approx&  \frac{1}{K} \sum_{k=1}^K I[ \tilde{C}(x_k)=G ]\, I[ C(x_k)=S ]\\
\mathbb{P}[C(x)=S] &\approx& \frac{1}{K} \sum_{k=1}^K I[ C(x_k)=S ]\\
\mathbb{P}[C(x)=G] &\approx& \frac{1}{K} \sum_{k=1}^K I[ C(x_k)=G ]
\end{eqnarray}

The relative error\footnote{By relative error, we mean the standard deviation of the Monte Carlo estimator divided by the mean value.} of these straightforward Monte Carlo estimators can be quite large however because the probability of snow $\mathbb{P}[C(x)=S]$ is small.  The central limit theorem implies that $K$ will therefore need to be incredibly large to obtain a Monte Carlo standard error that is sufficiently smaller than $\mathbb{P}[C(x)=S]$.

\subsubsection{Importance Sampling}
\label{subsubsec:imp_samp}

To overcome the need for an intractable number of samples $K$, we use importance sampling  \citep[e.g.][]{liu2001monte}, which is commonly used for rare event simulation  \citep[e.g.][]{denny2001introduction,rubino2009rare}. Instead of drawing $K$ samples of $x$ uniformly from the point cloud, we can draw the samples from an alternative distribution where the probability of a point being a wind-blown snow particle is larger, thus reducing the relative error of the Monte Carlo estimate. The probability of randomly selecting point $k$ under a uniform distribution over all points is $P_k = \frac{1}{K_{all}}$.  Assume we instead draw $K$ samples from a distribution over the points with probabilities $Q_k$. The Monte Carlo estimates of the required probabilities is then
\begin{eqnarray}
\mathbb{P}[(\tilde{C}(x) =S) \land (C(x)=G)] &\approx& \frac{1}{K} \sum_{k=1}^K I[ \tilde{C}(x_k)=S ]\, I[ C(x_k)=G ] \frac{P_k}{Q_k}\label{eqn:is1}\\
\mathbb{P}[(\tilde{C}(x) =G) \land (C(x)=S)] &\approx& \frac{1}{K} \sum_{k=1}^K I[ \tilde{C}(x_k)=G ]\, I[ C(x_k)=S ] \frac{P_k}{Q_k}\\
\mathbb{P}[C(x)=S] &\approx& \frac{1}{K} \sum_{k=1}^K I[ C(x_k)=S ] \frac{P_k}{Q_k}\\
\mathbb{P}[C(x)=G] &\approx& \frac{1}{K} \sum_{k=1}^K I[ C(x_k)=G ] \frac{P_k}{Q_k},\label{eqn:is4}
\end{eqnarray}
where the weight $P_k/Q_k$ is needed to account for the fact that the new samples $x_k$ stem from proposal distribution $Q$. This approach enables us to accurately estimate the false positive rate without needing to manually classify hundreds of thousands of samples. 

\subsubsection{Choosing Q}
\label{subsubsec:Q}

We apply a two-step process for drawing a sample $x_k\sim Q$ from the proposal distribution.  First, we decide whether the sample should be an $S$ sample or a $G$ sample (according to the approximate classifier), and then we uniformly select $x_k$ from the points in that class.  In this process we are free to choose the probability $\mathbb{P}[S]$ of initially choosing the snow class $S$.  Note that $\mathbb{P}[G]=1-\mathbb{P}[S]$.  The probability $\mathbb{P}[S]$ is therefore a parameter in our proposal distribution $Q$ that we can strategically choose to reduce the error in our Monte Carlo estimates. 

Mathematically, the probability of selecting a point $x_k$ is
\begin{equation}
\mathbb{P}[x_k] = \mathbb{P}[x_k | \tilde{C}(x_k)=S]\mathbb{P}[S] + \mathbb{P}[x_k | \tilde{C}(x_k)=G]\mathbb{P}[G].
\end{equation}
Assuming that the points are chosen uniformly within each class, we have 
\begin{equation}
\mathbb{P}[x_k] = \frac{1}{N_S}I[\tilde{C}(x_k)=S]\mathbb{P}[S] + \frac{1}{N-N_S}I[\tilde{C}(x_k)=G]\mathbb{P}[G],
\end{equation}
where $N$ is the total number of points in the point cloud and $N_S$ is the number of points satisfying $\tilde{C}(x)=S$. In our importance sampling approach, the goal is to strategically choose the proposal distribution to increase the likelihood of seeing snow samples.  To accomplish these, we set $\mathbb{P}[S] = q_s$ (e.g., $q_s=1/2$) in the proposal, which results in proposal probabilities
\begin{equation}
Q_k = \frac{1}{N_S}I[\tilde{C}(x_k)=S] q_s + \frac{1}{N-N_S}I[\tilde{C}(x_k)=G] (1-q_s).
\end{equation}
With this proposal, the importance weights in \eqref{eqn:is1}--\eqref{eqn:is4} are given by 
\begin{equation}
\frac{P_k}{Q_k} = \frac{1}{\frac{N}{N_S}I[\tilde{C}(x_k)=S] q_s + \frac{N}{N-N_S}I[\tilde{C}(x_k)=G] (1-q_s)}.
\end{equation}

Using our approximate classifier in this manner enables us to reduce the error in our Monte Carlo estimates of the false positive rate. Note that for rare events, this choice of $Q$ does not aid us in reducing the variance of the Monte Carlo estimate of the false negative rate. False negatives are events that our approximate classifier did not identify and, empirically, they occur very infrequently in this data. Thus we would require a better approximate classifier than the one we are trying to evaluate to generate a distribution $Q$ that aids us in estimating the false negative rates. For completeness, we present equations for estimating false negative rates.

\subsubsection{Manual Point Classifier}
\label{subsubsec:man_classifier}

\begin{figure}[H]
    \centering
    \includegraphics[width=13.5cm]{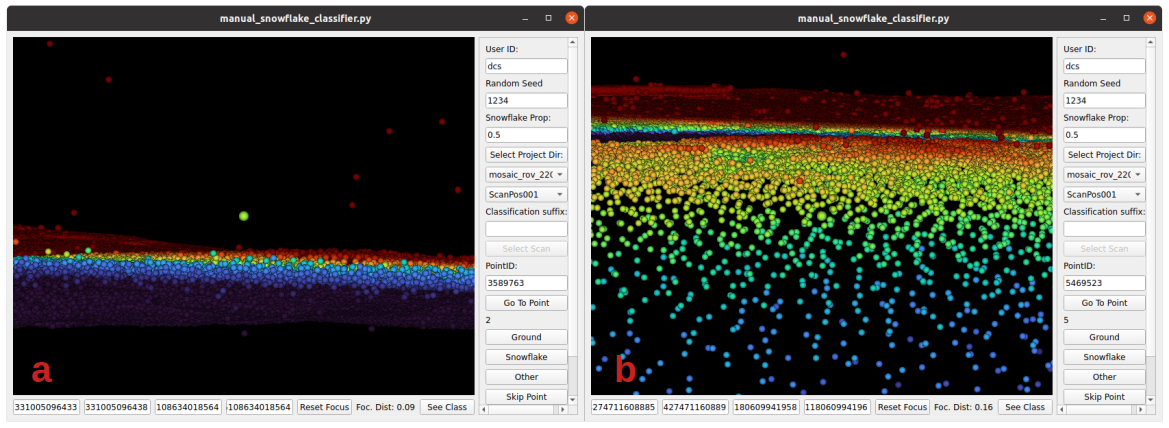}
    \caption{Screenshots of our visualization interface focused on a point that is a wind-blown snow particle (a) and a point on the snow surface (b). In both cases, the sampled point that the user is asked to classify is the green point in the center of visualization window, which is slightly larger than the other points. The points are colored by their vertical coordinate (blue is low, red is high) and the color window is 2 cm wide and centered on the height of the sampled point. With the mouse the user can rotate the scene around the sampled point and zoom in and out to identify whether the point is part of the surface or not. The user can either click the buttons on the righthand side of the interface to classify points or press `g' (for ground or snow surface) and `f' (for flake or snow particle) to process points efficiently.}
    \label{fig:gui}
\end{figure}

The importance sampling approach outlined above requires us to manually classify $K$ points drawn from the distribution $Q$. Due to saltation, most wind-blown snow particles are close to the ground and can be difficult to identify in conventional visualization software (e.g., Paraview, CloudCompare). Additionally, being able to automatically draw points from the distribution $Q$ will reduce the number of points $K$ needing manual classification. To address this, we developed an interactive visualization interface (see Fig. \ref{fig:gui} for example) that automatically samples points from $Q$ and displays them to the user. The color scale of the displayed points is set to aid the user in determining the elevation of the sampled point relative to its neighbors. Furthermore, the point is set as the focal point of the interactive visualization window. This enables the user to rotate the scene around the sampled point and exploit depth perception to identify whether it sits above the surface. For example, the sampled point in Fig. \ref{fig:gui}a is less than 1 cm above the snow surface, however this visualization interface makes it easy to identify as wind-blown snow particle. The sampled point in \ref{fig:gui}b is hard to distinguish from the neighboring points in this still image because it is part of snow surface. Rotating the scene makes it readily identifiable as such. The interface is implemented in Python \citep{rossum_python_2010} using VTK \citep{schroeder_will_visualization_2006} and is publicly available \citep{clemens-sewall_davidclemenssewallflake_out_2021}. With this tool, an experienced user can manually classify approximately 600 points per hour.

\section{Results}
\label{sec:Results}

We applied the FlakeOut filter (Section \ref{subsec:Filter}), early return filtering, and radius outlier removal (using optimized implementation from \citealt{zhou_open3d_2018}) to the nine TLS scan positions described in Section \ref{subsec:Data} (Table \ref{table:counts}). Each scan position contained approximately 16.5 million points. For radius outlier removal, we used the parameters $radius=0.14$ m and $nb\_points=4$ \citep{zhou_open3d_2018}. See Section \ref{sec:Discussion} for discussion about radius outlier removal parameters. We attempted to apply statistical outlier removal as well, however, it was not computationally feasible to apply to such large point clouds. 

\begin{table}[H]
\centering
 \begin{tabular}{|c c c c|} 
 \hline
 Filter & Number Filtered & Fraction Filtered (/1000) & Runtime \\ [0.5ex] 
 \hline
 FlakeOut & \num{1.9e4} & 1.2 & 96 s\\
 Early Return & \num{6.3e4} & 3.8 & 3 s\\
 Radius Outlier & \num{7.1e5} & 43 & 425 s\\
 \hline
 \end{tabular}
\caption{Comparison of mean (across all nine scan positions) number of points filtered, mean fraction filtered per thousand points and runtimes for proposed filter (FlakeOut) and other filters. Runtimes are for a single scan position and are on an Intel(R) Xeon(R) E-2176M  CPU @ 2.70GHz. The Radius Outlier filter is multithreaded \citep{zhou_open3d_2018} and was run using all 6 cores. FlakeOut and Early Return filters are not multithreaded.}
\label{table:counts}
\end{table}

For each scan position and each filter, we manually classified at least 100 points using the manual point classifier described in section \ref{subsubsec:man_classifier} with $q_s=0.5$ ($q_s$ defined in Section \ref{subsubsec:Q}). We estimated false positive according to Section \ref{subsec:Validation} (Table \ref{table:results}). The false positive rate of FlakeOut is an order of magnitude less than the next best filter. For the reasons described in Section \ref{subsubsec:imp_samp}, the variance of the Monte Carlo estimates of the false negative rates are sufficiently high that the 95\% confidence intervals would span the entire possible range. Hence we do not show them. Out of the manually classified points, FlakeOut had half as many false negatives as the next best filter (Table \ref{table:negatives}). However, we should stress that the total number of points is so few that we cannot make a statistical comparison.

\begin{table}[H]
\centering
 \begin{tabular}{|c c c|} 
 \hline
 Filter & False Positive Rate & FPR 95\% CI \\ [0.5ex] 
 \hline
 FlakeOut & \num{2.8e-4} & \num{2.2} - \num{3.2e-4} \\
 Early Return & \num{3.7e-3} & \num{3.3} - \num{4.0e-3} \\
 Radius Outlier & \num{4.2e-2} & \num{3.8} - \num{4.6e-2} \\
 \hline
 \end{tabular}
\caption{Comparison of mean false positive rates, and confidence intervals for proposed filter (FlakeOut) and other filters.}
\label{table:results}
\end{table}

\begin{table}[H]
\centering
 \begin{tabular}{|c c c |} 
 \hline
 Filter & True Negatives & False Negatives \\ [0.5ex] 
 \hline
 FlakeOut & 539 & 2 \\
 Early Return & 451 & 5 \\
 Radius Outlier & 490 & 4 \\
 \hline
 \end{tabular}
\caption{Comparison of the number of true negatives and false negatives for each filter.}
\label{table:negatives}
\end{table}

\section{Discussion}
\label{sec:Discussion}

The FlakeOut filter outperforms early return filtering and radius outlier removal at filtering wind-blown snow particles from TLS data on sea ice in terms of false positive rates (Table \ref{table:results}). This is because neither early return filtering nor radius outlier removal account for the scanning geometry of TLS on generally flat surfaces like sea ice and the characteristics of wind-blown snow. Below, we discuss how accounting for these factors enables FlakeOut to improve on these two common filtering methods in this context.

The shallow scan angle in TLS data (most points are near the horizon and most incidence angles are near 0\degree) causes early return filtering to miss snow particles for two reasons. First, any snow particles above the horizon will produce last returns (there's no surface behind these snow particles to generate another return). Second, manual inspection reveals that many airborne snow particles near the surface also generate last returns. One reason for this may be that mean particle diameter in the blowing snow layer increases towards the surface \citep{schmidt_properties_1982,nishimura_blowing_2005}. The laser beam diameter is substantially larger than large blowing snow particles (e.g. for the VZ-1000 1.5 mm at 5 m range whereas \citet{nishimura_blowing_2005} found blowing snow particles up to 450 $\mu$m in diameter). However, the attenuation from these larger particle combined with the oblique incidence angle may reduce the intensity of the surface return to below the detector's limit. The vertical z-score filter (Section \ref{subsubsec:z_score}) addresses both of these issues. In contrast, early return filtering mistakenly removes many points on the snow and ice surface---leading to a false positive rate that is an order of magnitude greater than the false positive rate of FlakeOut. These early returns from the surface are produced by oblique incidence angles and surface roughness. The visible region filter (Section \ref{subsubsec:vis_region}) addresses this problem by only filtering early returns if they are clearly not part of the snow and ice surface.

Radius outlier removal poorly filters blowing snow, because the TLS scanning geometry causes the point spacing to increase with distance from the scanner, shadowing and slope aspect creates variable point density of the surface, and most blowing snow particles are near the surface. Thus, even for a perfectly flat surface, any fixed set of parameters will incorrectly filter returns from the surface far away from the scanner, and incorrectly fail to filter snow particles that are close to the surface. The parameters used in this paper were chosen such that on a flat surface, they would only incorrectly filter any returns greater than 150 m of the scanner. Note that a cursory inspection of the output of radius outlier removal reveals that it correctly identifies most obvious snow particles––those tens of centimeters above the surface. However, these obvious snow particles make up only a small proportion of the wind-blown snow particles. Most blowing snow particles are in the saltating layer close to the snow surface. In fact, out of the 476 manually classified points that radius outlier removal had identified as wind-blown snow particles, not a single one was manually classified as such. Manual inspection of the results indicates that radius outlier removal is primarily filtering points far from the scanner and areas where shadowing leads to lower point densities. Dynamic Radius Outlier Removal \citep{charron_-noising_2018} addresses point spreading with distance, but not variable point density caused by shadowing and slope aspect. In fact, purely distance-based filters (DROR, radius outlier removal, and statistical outlier removal) will all incorrectly filter surface points in areas of low point density caused by shadowing and slope aspect. Although statistical outlier removal was not computationally feasible on the large point clouds used in this data set, it uses a single distance parameter for the entire point cloud and hence suffers the same point density problems as radius outlier removal.

FlakeOut improves on distance-based filters in three ways. First, the visible region filter explicitly accounts for shadowing by assessing the boundaries of the visible region from the scanner's perspective. Second, the vertical z-score incorporates the core idea of the DROR filter---that filter thresholds should vary locally throughout the scanned area---and improves on it by defining local regions by fixed numbers of points and thresholding only based on the vertical coordinate. Using a set number of points, instead of a set radius (radius outlier removal) or a radius that varies only with distance from the scanner (DROR), allows the local regions to vary in size and accounts for shadowing. Thresholding only in the vertical coordinate prevents us from removing points that are from the surface and are just distant from other surface points due to distance from the scanner and/or shadowing. FlakeOut is more than four times faster than radius outlier removal (Table \ref{table:counts}) and could be optimized to operate faster still.

\section{Conclusions}
\label{sec:Conclusion}

In this paper, we present FlakeOut, the first filter designed to remove wind-blown snow particles from Terrestrial Laser Scanning data. Wind-blown snow particles are small, concentrated near the snow surface, and commonly present in tundra, prairie, ice sheet, and sea ice environments. FlakeOut accounts for TLS scanning geometry and the characteristics of wind-blown snow particles to filter snow particles while minimizing the amount of the surface points it incorrectly removes. FlakeOut achieves a false positive rate of \num{2.8e-4}, an order of magnitude lower than standard techniques for removing airborne particles. Additionally, we provide mathematical techniques for efficiently estimating accuracy metrics of filters applied to rare events like wind-blown snow particles in TLS data.

\section{Acknowledgements} 
Data used in this manuscript were produced as part of the international Multidisciplinary drifting Observatory for the Study of the Arctic Climate (MOSAiC) with the tag MOSAiC20192020 and the Project\_ID: AWI\_PS122\_00. We thank all people involved in the expedition of the Research Vessel Polarstern \citep{knustpolarstern2017} during MOSAiC in 2019-2020 as listed in Nixdorf et al. (2021).

\section{Funding information} 
D.C.-S., D.P., C.P., and I.R. conducted this work under NSF OPP-1724540. M.P. conducted this work under Office of Naval Research grant \#N00014-20-1-2595.

\section{Competing interests} 
The authors declare that they have no known competing financial interests or personal relationships that could have appeared to influence the work reported in this paper.

\section{Data accessibility statement} 
The data used in this work will be accessible at the Arctic Data Center (has been submitted, will include citation and doi when available). 

\bibliographystyle{elsarticle-harv} 
\bibliography{flake_out}





\end{document}